\documentclass{epl}
\usepackage{graphicx,rotating,subfigure,amsmath,amsfonts,amssymb,delarray,cite}
\renewcommand{\vec}[1]{\boldsymbol #1}

\newcommand{\im}{\text{i}}
\title{Consequences of spin-orbit coupling for the Bose-Einstein condensation of magnons} 
\shorttitle{Consequences of spin-orbit coupling etc.} 
\author{J. Sirker \and A. Wei{\ss}e \and O.P. Sushkov}
\institute{School of Physics, The University of New South Wales,
  Sydney 2052, Australia
}
\begin{document}
\maketitle
\begin{abstract}
  In the first part we discuss how the BEC picture for magnons is modified by
  anisotropies induced by spin-orbit coupling. In particular we focus on the
  effects of antisymmetric spin interactions and/or a staggered component of
  the $g$ (gyromagnetic) tensor. Such terms lead to a gapped quasiparticle
  spectrum and a nonzero condensate density for all temperatures so that no
  phase transition occurs. We contrast this to the effect of crystal field
  anisotropies which are also induced by spin-orbit coupling. In the second
  part we study the field-induced magnetic ordering in TlCuCl$_3$ on a
  quantitative level. We show that the usual BEC picture does not allow for a
  good description of the experimental magnetisation data and argue that
  antisymmetric spin interactions and/or a staggered $g$ tensor component are
  still crucial, although both are expected to be tiny in this compound due to
  crystal symmetries. Including this type of interaction we obtain excellent
  agreement with experimental data.
 \end{abstract}
\section{Introduction}
It has been argued that the phase transition at a critical applied magnetic
field in certain spin systems with an excitation gap $\Delta$ as for example
integer-spin antiferromagnetic chains \cite{AffleckBEC}, weakly coupled
two-leg ladders \cite{GiamarchiTsvelik} or three-dimensional dimer
systems \cite{NikuniOshikawa,RiceScience}
may be regarded as a Bose condensation. In these systems the lowest excited
state is a triplet of massive bosons. A magnetic field $H$ causes a Zeeman
splitting of the triplet with the lowest mode crossing the ground state at a
critical field $H_c = \Delta/g\mu_B$. The ground state for $H>H_c$ then
becomes a BEC of this low-energy boson. The density $n$ of the boson mode is
directly related to the magnetisation per site $m = g\mu_B n$. In principle,
spin-gap compounds offer therefore the exciting possibility to study BEC in a
system where the density $n$ is tunable by the external magnetic field which
acts as a chemical potential for the bosons. However, as the triplet
excitations (magnons) interact via a hard-core repulsion a description as a
weakly interacting Bose gas is only meaningful if the average distance between
the magnons $l\sim n^{-1/3}$ is much larger than the s-wave scattering length
$a$ which is the characteristic length scale representing the influence of the
repulsive potential. This implies that $a/l\sim n^{1/3}a\ll 1$ so that the
magnons have to be dilute. In this case the well-established gas approximation
\cite{FetterWalecka} which involves a systematic expansion in terms of the
small parameter $n^{1/3}a$ is applicable and even the finite temperature
properties of the interacting Bose gas can be studied analytically
\cite{GriffinRep}.

A dilute magnon gas is realized in the spin dimer system TlCuCl$_3$ in
magnetic fields $H\sim 6-7$ T and considerable interest has focused onto this
compound within recent years
\cite{RiceScience,NikuniOshikawa,MatsumotoNormandPRL,CavadiniHeigold,ChoiGuentherodt,ShermanLemmens}.
TlCuCl$_3$ has an excitation gap $\Delta \approx 0.7$~meV in zero magnetic
field and a bandwidth $W\sim 6.3$~meV \cite{CavadiniHeigold}. The dimers in
this compound are formed by the $S=1/2$ spins of the Cu$^{2+}$ ions and
superexchange interactions are mediated by the Cl$^-$ ions. The crystal
structure can be considered as two-leg ladders formed by these dimers.
However, inelastic neutron scattering (INS) \cite{CavadiniHeigold} has
revealed that
the magnons show a considerable dispersion in all 
spatial directions indicating that the interladder interactions are strong.
TlCuCl$_3$ therefore has to be considered as a three-dimensional (3D)
interacting dimer system. For fields $H\gtrsim H_c\sim 5.6$~T a long-range
magnetic ordering below some critical temperature $T_c$ has been detected.
It has been shown that this transition and the overall shape of the
magnetisation curves as a function of temperature can be qualitatively
described as the BEC of magnons \cite{NikuniOshikawa}. Recently, the
excitations for $H>H_c$ have been measured by INS and the lowest mode has been
interpreted as the gapless Goldstone mode characteristic for a Bose condensed
phase \cite{RueggNature}.

In the first part of this letter we will show how to include antisymmetric
spin interactions or a staggered component of the $g$ tensor into a
Hartree-Fock-Popov (HFP) treatment \cite{GriffinRep,NikuniOshikawa} of an
interacting dilute Bose gas. Such terms will in general break the axial $U(1)$
symmetry of the system so that a Goldstone mode no longer exists. Furthermore
the condensate density becomes nonzero for all temperatures so that no phase
transition occurs. These findings are in agreement with \cite{MitraHalperin}
where the same type of anisotropy in a Haldane spin chain has been considered.
We will discuss the differences to the effect of crystal field anisotropies
which can also break $U(1)$ symmetry and affect Bose condensation
\cite{AffleckBEC}. Our analysis will allow us to investigate the effects of
such anisotropies on the magnon density in 3D dimer systems at finite
temperatures quantitatively and we present a detailed study of the
field-induced magnetic ordering in TlCuCl$_3$ along these lines in the second
part of this letter.
\section{General scenario}
\begin{figure}[!htp]
\onefigure[width=0.7\columnwidth]{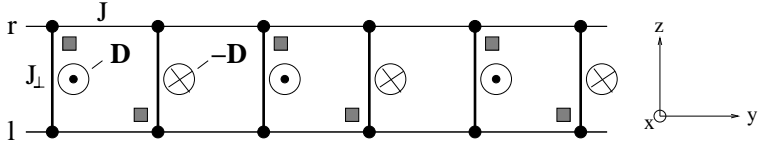}
\caption{Ladder with rung couplings $J_\perp$ and $J\ll J_\perp$ along the legs. $\vec{D}$ is staggered along the x-axis. The
  squares denote some environment preventing inversion centres within each dimer.}
\label{fig_ladder}
\end{figure}

We want to restrict ourselves here to 3D spin-1/2 systems where the spin gap
is due to some kind of explicit dimerisation.
A useful approach to describe the excitations in 
such systems is the bond operator representation for spins introduced in
\cite{ChubukovSachdevBhatt}. The starting point is the strong coupling
ground state $|s\rangle$ where each dimer at site $i$ is in singlet
configuration $|i,s\rangle$. It is then natural to introduce operators
$t^\dagger_{i\alpha}$ which create local triplet excitations $|i,\alpha\rangle
= t^\dagger_{i\alpha}|i,s\rangle$
with $|i,+\rangle = -|\uparrow\uparrow\rangle$, $|i,-\rangle =
|\downarrow\downarrow\rangle$ and $|i,0\rangle =
(|\uparrow\downarrow\rangle+|\downarrow\uparrow\rangle)/\sqrt{2}$. Due to
hopping between the dimers the three triplet components acquire a dispersion
$\Omega_{\vec{k}\alpha} = \Omega_{\vec{k}0}-\alpha g\mu_B H$ with a band
minimum $\Omega_{\vec{q}_0\alpha}=\Delta-\alpha g\mu_B H$ at momentum
$\vec{q}_0$ depending on microscopic details of the exchange interactions. The
triplets are subject to a hard-core constraint which can be taken into account
by introducing an infinite on-site repulsion
\begin{equation}
\label{eq4}
\mathcal{H}_U = U\sum_{i,\alpha,\beta}
t^\dagger_{i\alpha}t^\dagger_{i\beta}t_{i\alpha}t_{i\beta}, \quad
U\rightarrow\infty
\end{equation}
with $\alpha,\beta = -,0,+$. Note that the renormalisation of the triplet
dispersion $\Omega_{\vec{k}\alpha}$ due to the two-particle scattering vertex
$v(\vec{k},\omega)$ corresponding to the interaction (\ref{eq4}) can be
calculated exactly in the dilute limit by a summation of ladder diagrams
\cite{KotovSushkov}. For magnetic fields $H\gtrsim H_c$ and temperatures
$T<\Delta$ it is sufficient to take only the lowest triplet mode ($\alpha=+$)
into account. In this case only particles near the band minimum at $\vec{q}_0$
are excited so that the generally energy and momentum dependent
$v(\vec{k},\omega)$ can be replaced by a constant $v_0=v(\vec{q}_0,0)$. With
the definitions $\epsilon_{\vec{k}}\equiv\Omega_{\vec{k}0}-\Delta$ and
$t_{\vec{k}}\equiv t_{\vec{k}+}$ the Hamiltonian for the lowest mode is given
by
\begin{equation}
\label{eq6}  
\mathcal{H} = \sum_{\vec{k}} \left(\epsilon_{\vec{k}}-\mu_0\right)t_{\vec{k}}^\dagger t_{\vec{k}}
+\frac{v_0}{2}\sum_{\vec{k},\vec{k}',\vec{q}} t_{\vec{k}+\vec{q}}^\dagger t_{\vec{k}'-\vec{q}}^\dagger
t_{\vec{k}} t_{\vec{k}'} 
\end{equation}
where $\mu_0 = g\mu_B (H-H_c)$. Here we want to consider 
a perturbation linear in $t$, $t^\dagger$ 
\begin{equation}
\label{eq10}  
\mathcal{H}' = \im\gamma (t_{\vec{q_0}} - t^\dagger_{\vec{q_0}}) 
\end{equation}
where $\gamma$ is a small parameter. This term respects parity and time
reversal symmetry and will therefore be non-zero in general if not forbidden
by additional crystal symmetries. Clearly its effect will be {\it
  non-perturbative} at fields $H\sim H_c$ because it mixes the singlet ground
state and the triplet soft mode. Physically such a term can originate from
{\it Dzyaloshinsky-Moriya (DM) interactions}
$\sim\vec{D}\cdot\left(\vec{S}_i\times\vec{S}_j\right)$ between spins at sites
$i,j$ where $\vec{D}$ is the DM vector. For illustration we consider the
two-leg spin-ladder shown in fig.~\ref{fig_ladder}.
In the limit $J_\perp\gg J$
the triplet dispersion for this model is given by
$\Omega_{k\alpha}=J_\perp +J\cos k-\alpha g\mu_B H$ and has a minimum at
$q_0=\pi$. Consider now the additional antisymmetric exchange
\begin{equation}
\label{DM}
\mathcal{H}_{DM} = D_x\sum_i (-1)^i \left(S^l_{iy} S^r_{iz} - S^l_{iz} S^r_{iy}\right)
\end{equation}
where indices $l,r$ label the two spins within the dimer as shown in
fig.~\ref{fig_ladder}. Using the identity \cite{ChubukovSachdevBhatt}
$ S^{l,r}_\alpha =(\pm t_\alpha \pm
t_\alpha^\dagger-\im\vec{\epsilon}_{\alpha\beta\gamma} t_\beta^\dagger
t_\gamma )/2$ with $\alpha = x,y,z$ this is easily transformed into
$\mathcal{H}_{DM}=\im D_x\sum_i (-1)^i (t_{ix} - t_{ix}^\dagger)/2$. Using
spiral instead of cartesian indices we find in momentum space
$\mathcal{H}_{DM}\sim\im D_x (t_\pi - t_\pi^\dagger)$ where $t$ denotes now
the lowest triplet mode ($\alpha=+$) as in (\ref{eq6}). This is exactly of the
proposed form (\ref{eq10}) with the triplet operators $t_\pi^{(\dagger)}$
acting at the band minimum. It can be shown by an exact transformation that
the considered DM interaction produces an effective staggered field if a
uniform magnetic field is applied \cite{OshikawaAffleck}. A staggered magnetic
field can also originate directly from a staggered $g$ tensor
\cite{MitraHalperin}. Both effects are induced by spin-orbit coupling and can
contribute to (\ref{eq10}).  Note, however, that the interaction (\ref{eq10})
{\it itself} is forbidden by symmetry if there is an inversion center in the
middle of each dimer irrespective of the origin of this term.

Next we diagonalize (\ref{eq6}) with the perturbation (\ref{eq10}) included,
treating the interaction between non-condensed magnons in the one-loop (HFP)
approximation.
We want to emphasize again that taking only these diagrams into account is not
an uncontrolled approximation but instead the first order in a systematic
expansion in the gas parameter $n$. 
First, we introduce new operators $c_{\vec{k}}$ by $t_{\vec{k}} = c_{\vec{k}}
+\im\delta_{\vec{k},\vec{q}_0}\eta$ where $\eta$ is a real number.  The
density of condensed magnons per dimer $n_0$ is then given by $n_0=\eta^2$.
Ignoring a momentum independent term, the Hamiltonian splits into 2 parts
$\mathcal{H}=\mathcal{H}_{\text{lin}}+\mathcal{H}_{\text{bilin}}$
with
$\mathcal{H}_{\text{lin}} = \im(2\tilde{n}v_0\eta+v_0\eta^3-\mu_0\eta-\gamma)(c_{\vec{q}_0}^\dagger -c_{\vec{q}_0})$
where $\tilde{n}$ denotes the density of non-condensed magnons per dimer
and
\begin{equation}
\label{eq13} 
\mathcal{H}_{\text{bilin}} = \sum_{\vec{k}} \{ \mathcal{A}_{\vec{k}} c_{\vec{k}}^\dagger c_{\vec{k}} - \Sigma_{12}(c_{\vec{k}}^\dagger
    c_{-{\vec{k}}}^\dagger + h.c.)/2 \} \; .
\end{equation}
Here $\mathcal{A}_{\vec{k}} = \epsilon_{\vec{k}} - \mu_0+\Sigma_{11}$ with the normal
self-energy $\Sigma_{11} = 2v_0\tilde{n}+2v_0\eta^2$ whereas the anomalous
self-energy is given by $\Sigma_{12} = v_0\eta^2$. By a Bogoliubov
transformation we find the quasiparticle spectrum $E_{\vec{k}} =
(\mathcal{A}_{\vec{k}}^2-\Sigma_{12}^2)^{1/2}$.  Additionally we have to demand that
$\mathcal{H}_{\text{lin}}$ vanishes
\begin{equation}
\label{eq14} 
-\mu_0\eta -\gamma + \eta\left(\Sigma_{11}-\Sigma_{12}\right) = 0 \; .
\end{equation}
To stress the differences between BEC with and without $\mathcal{H}'$ we first briefly
summarise the case $\gamma = 0$ which is the usual BEC scenario for a weakly
interacting Bose gas in the HFP approximation.
In this case we have to distinguish further between the case $\eta=0$, i.e., no
magnons are condensed and the case with nonzero condensate density $\eta\neq
0$. 
For $\eta = 0$ the subcondition (\ref{eq14}) is identically fulfilled, the
anomalous self-energy $\Sigma_{12}$ vanishes and the quasiparticle spectrum is
identical to the bare triplet spectrum with an effective chemical potential
$\mu = \mu_0 -2v_0\tilde{n}$.
The density of magnons in this phase is therefore given by the usual Bose
distribution
$\tilde{n} = (1/N)\sum_{\vec{k}}[\exp
(\beta(\epsilon_{\vec{k}}-\mu))-1]^{-1}$. BEC occurs when the chemical
potential $\mu$ vanishes so that the density $n_c$ at the critical point is
given by $n_c = \mu_0/2v_0$. At temperatures below the critical point
($T<T_c$) $\eta$ will be nonzero and the condition (\ref{eq14}) for $\gamma =
0$ becomes equivalent to the {\it Hugenholtz-Pines} theorem \cite{GriffinRep}
$\mu_0=\Sigma_{11}-\Sigma_{12}$ and guarantees the existence of a Goldstone
mode, i.e., the quasiparticle spectrum $E_{\vec{k}} =
\sqrt{\epsilon_{\vec{k}}^2+2\epsilon_{\vec{k}}n_0v_0}$ is gapless. The number
of non-condensed magnons $\tilde{n}$ in this phase is given by
\begin{equation}
\label{eq8}  
\tilde{n} = -\frac{1}{2}+\frac{1}{N}\sum_{\vec{k}}
\frac{\epsilon_{\vec{k}}+n_0v_0}{2E_{\vec{k}}}\coth\left(\frac{\beta E_{\vec{k}}}{2}\right)
\end{equation}
whereas the number of condensed magnons can be determined from
eq.~(\ref{eq14}) which has to be solved self-consistently together with
(\ref{eq8}).

Now we want to compare this with the case $\gamma\neq 0$.
From eq.~(\ref{eq14}) it follows immediately that this changes the situation
qualitatively {\it irrespective of the magnitude of $\gamma$} because $\eta$
must be nonzero for all temperatures so that there will be always condensed
magnons and no phase transition will occur. Furthermore there is no longer a
Hugenholtz-Pines theorem and consequently the quasiparticle spectrum
\begin{equation}
\label{eq15} 
E_{\vec{k}}=\sqrt{\left(\epsilon_{\vec{k}}+|\gamma|/\sqrt{n_0}\right)^2+2\left(\epsilon_{\vec{k}}+|\gamma|/\sqrt{n_0}\right)v_0n_0}
\end{equation}
is gapped. This is expected because the component of the DM interaction
perpendicular to the applied field $H$ breaks the $U(1)$ symmetry so that
there is no longer a Goldstone mode. Using the spectrum (\ref{eq15}), the
non-condensed magnon density $\tilde{n}$ can be again calculated by
eq.~(\ref{eq8}) with $\epsilon_{\vec{k}}$ being replaced by
$\epsilon_{\vec{k}}+|\gamma|/\sqrt{n_0}$. The condensed density is determined
by eq.~(\ref{eq14}) so that there are again two equations which have to be
solved self-consistently.

Finally we want to discuss the difference between the perturbation
(\ref{eq10}) originating from DM interactions or a staggered $g$ tensor and
single-ion anisotropies $\sim D(S^z_i)^2+E[(S^x_i)^2-(S^y_i)^2]$. (Exchange
anisotropies
have the same effect as single-ion anisotropies) In terms of triplet operators
a single-ion anisotropy produces a perturbation {\it bilinear} in the triplet
operators $\mathcal{H}_{\text{pert}}=\tilde{\gamma}(t_kt_{-k}+h.c.$) which
results from the $E$-term provided that the magnetic field is along the
$z$-axis. Here $\tilde{\gamma}$ is again a small parameter and we assume
$\tilde{\gamma}>0$ without loss of generality. Consequently the condition for
the vanishing of the linear terms now becomes $-\mu_0\eta -2\tilde{\gamma}\eta
+ \eta\left(\Sigma_{11}-\Sigma_{12}\right) = 0$ and
$\tilde{\gamma}\neq 0$ does not imply $\eta\neq 0$ as before. Therefore a
phase transition will still occur and $n_0=0$ for $T>T_c$. The quasiparticle
gap will vanish exactly at the critical point, however, it will reopen again
below $T_c$. Additionally, the critical point itself will be slightly shifted.
Although in both cases the quasiparticle spectrum is gapped for $n_0\neq 0$,
the gap only weakly depends on the condensate density for the single-ion
anisotropy case contrary to what was found before (see eq.~(\ref{eq15})).
\section{TlCuCl$_3$}
\begin{figure}[!t]
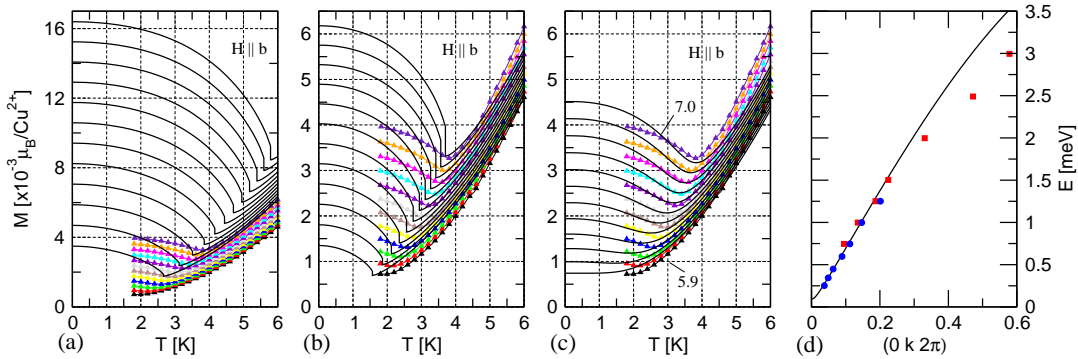

\onefigure[width=1.0\columnwidth]{v0_Delta_0.72_all_combined5.eps}
\caption{Experimental magnetisation curves (symbols) for
  different magnetic fields $H=5.9,6.0,6.1,\cdots,7$~T taken from
  ref.~\cite{NikuniOshikawa} compared to the theoretically calculated (solid
  lines) with $\Delta = 0.67$~meV where (a) $v_0 = 9.8$~meV and (b) $v_0 =
  25$~meV. In (c) the result with (\ref{eq10}) included is shown where
  $\gamma=10^{-3}$~meV, $\Delta = 0.72$~meV and $v_0 = 27$~meV.  (d)
  Quasiparticle spectrum (\ref{eq15}) (solid line) for $H=14$~T ($n_0\approx
  2.3\cdot 10^{-2}$ at $T\lesssim 1.5$~K) with all other parameters as in (c).
  INS data for $T=50$~mK (circles) and $T=1.5$~K (squares) taken from
  ref.~\cite{RueggNature} are shown for comparison.}
\label{fig1}
\end{figure}

For TlCuCl$_3$ detailed magnetisation measurements have been performed for
magnetic fields $H\gtrsim H_c$ with typical magnon densities $n\sim 10^{-3}$
\cite{NikuniOshikawa} satisfying the condition of diluteness. This compound is
therefore particularly suited to test the predictions of the BEC theory within
the HFP approximation even quantitatively. This is the purpose of the
remainder of this letter. TlCuCl$_3$ crystallises in the space group $P2_1/c$
with two dimers per unit cell formed by the $S=1/2$ spins of the Cu$^{2+}$
ions \cite{CavadiniHeigold}.  Because the centre of each dimer is an inversion
centre, the interaction (\ref{eq10}) is forbidden by symmetry making,
apparently, a description as a usual BEC valid. To do so it is not sufficient
to use a magnon dispersion of the form $\Delta+\vec{k}^2/2m$
\cite{NikuniOshikawa}. Although a quadratic dispersion is indeed expected
close to the band minimum its applicability is restricted to small excitation
energies and we find that this simplification here is only justified for
temperatures $T<1$~K which is well below the experimental temperature range
\footnote{The experimentally observed deviation from the universal BEC power
  law $n_c\sim T_c^\alpha$ with $\alpha=3/2$ is therefore not astonishing
  because this exponent is a direct consequence of a quadratic dispersion.}.
Matsumoto {\it et al.} \cite{MatsumotoNormandPRL} have successfully applied
the afore mentioned bond-operator technique to describe the real triplet
dispersion in TlCuCl$_3$. 
However, we have found that the mean-field treatment of the hard-core
constraint (\ref{eq4}) used in \cite{MatsumotoNormandPRL} is not sufficient
for a quantitative comparison with the magnetisation measurements in
\cite{NikuniOshikawa}. We therefore have calculated the renormalisation of the
dispersion due to the constraint in a systematic way by a summation of ladder
diagrams \cite{KotovSushkov}. Doing so we find a considerable renormalisation
of the superexchange parameters compared to ref.~\cite{MatsumotoNormandPRL}.
In particular this treatment allows us to calculate the scattering amplitude
$v_0$ directly and we find $v_0=1.6\, W=9.8$ meV \cite{Details}.
Additionally we note that the gap cannot be determined very accurately from
the INS data so that we have used $H_c=5.6$ T and $g=2.06$
\cite{GlazkovSmirnov} which yield $\Delta=g\mu_B H_c=0.67$ meV.  This value
was used as a constraint when calculating the renormalised dispersion. Having
the renormalised spectrum and $v_0$ at hand we can calculate the magnon
density and therefore the magnetisation as a function of temperature within
the standard BEC picture. As we are interested in temperatures $T\ll v_0$ we
can treat $v_0$ as temperature independent. The result is shown in
fig.~\ref{fig1}a.
Quite obviously the calculated magnetisation curves do not agree with the
experimental data. We want to point out that although the jump in the
magnetisation at the transition point is known to be an artefact of the HFP
treatment, this approximation is reliable for a dilute Bose gas apart from the
small temperature interval $|T-T_c|\lesssim n^{1/3}aT_c$ \cite{GriffinRep}
where we find $n^{1/3}a\lesssim 0.1$. Especially the huge overestimation of
the magnetisation at $T=0$ is definitely not due to an invalid approximation.

Taking $v_0$ as a fitting parameter we can obtain good agreement with
experiment up to the minima of the magnetisation curves with $v_0= 25$ meV
yielding the best result shown in fig.~\ref{fig1}b. A renormalisation of $v_0$
up to this value is by no means implausible as our calculation of $v_0$ takes
only magnon-magnon interactions (\ref{eq4}) into account.  Raman spectroscopy,
however, has revealed that low-energy optical phonons exist in TlCuCl$_3$
which interact with the spin system \cite{ChoiGuentherodt}. As a consequence a
reduction of the bare magnon bandwidth $W_0$ due to polaronic effects is
expected so that $v_0$, which is basically given by $W_0$, might be in fact
much larger. A reduction of the bandwidth by a factor 2-3 is not exceptional
and known in polaron physics for a long time \cite{Holstein}.  However, even
when using $v_0= 1.6\, W_0 = 25$ meV the theoretically calculated magnetisations
are still about 50\% too large at $T=0$.  More important, the standard BEC
scenario always predicts a sharp increase of the magnetisation below $T_c$
which {\it qualitatively} disagrees with experiment where only a slight and
smooth upturn is visible.

In recent ESR measurements \cite{GlazkovSmirnov} a direct singlet-triplet
transition has been observed which would be usually forbidden by spin
conservation. This yields some indication that a small perturbation of the
form (\ref{eq10}) is still present. The reason might be small static
distortions which violate the exact inversion symmetry within a dimer so that
$\mathcal{H}'$, although certainly tiny, becomes nonzero. 
We therefore tried to fit the measured magnetisation curves using the outlined
BEC theory with $\mathcal{H}'$ included. Because any magnetic field will then cause
a finite magnetisation, $H_c$ is no longer well defined.  However, fixing the
gap in a range consistent with the INS data we can still use
$H_c=\Delta/g\mu_B$ as a formal definition. With $\Delta = 0.72$~meV,
$v_0=27$~meV and a DM interaction $\gamma=10^{-3}$~meV we can obtain excellent
agreement with experiment as shown in fig.~\ref{fig1}c.
Most important, even this extremely small perturbation yields smooth minima
and a slow increase of the magnetisation at temperatures below the minima
consistent with experiment.
If $\gamma$ is indeed nonzero we expect a gapped quasiparticle spectrum.
In fig.~\ref{fig1}d we therefore compare the spectrum (\ref{eq15}) with the
INS data taken from ref.~\cite{RueggNature}. At low excitation energies our
result is in perfect agreement with these data. The deviations at higher
energies are expected because our approximation
$v(\vec{k},\omega)=v(\vec{q}_0,0)$ in eq.~(\ref{eq6}) is then no longer
justified. The predicted gap $\Delta\sim 0.1$ meV is only a factor 2 smaller
than the lowest measured excitation energies making it perhaps accessible in
future studies. We also mention that in systems where DM interactions are
allowed by symmetry they are typically of the order $\gamma\sim J(g-2)/g$
where $J$ is the isotropic superexchange \cite{Moriya}. For a DM interaction
within a TlCuCl$_3$ dimer this estimate yields $\gamma\sim 0.2$~meV which is
two orders of magnitude larger than the term considered here supporting our
statement that tiny violations of inversion symmetry are sufficient. However,
a nonzero anisotropy term raises questions about the orientation of the DM
vector $\vec{D}$. If $\vec{D}$ would be oriented along a specific axis
throughout the crystal as in our example in fig.~\ref{fig_ladder} it would be
possible to restore $U(1)$ symmetry by applying the magnetic field along the
same axis. In this configuration a sharp phase transition would be visible.
However, such dependencies on field direction have not been reported for
TlCuCl$_3$. Another possibility would be that small static distortions lead to
domains with different orientations of $\vec{D}$. In this case a component
perpendicular to $H$ could exist for each field direction. Provided that
indeed no anisotropy with respect to the field direction exists we believe
this is the most probable scenario taking the tininess of the needed DM term
into account. In principle it is also allowed by symmetry to construct a DM
term from the magnetic field itself, i.e.~it is possible that the magnetic
field induces the anisotropy and therefore determines the orientation of
$\vec{D}$. Finally we want to discuss a single-ion or exchange anisotropy term
as an alternative possible perturbation. As already mentioned before a sharp
phase transition will still occur with such a term included. Additionally, the
quasiparticle gap in the condensed phase will be almost independent of $n_0$
and small for a small perturbation. We therefore do not expect any qualitative
change for the magnetisation curves. Explicit calculations indeed show that
such a term basically yields a small shift of the critical temperature leaving
the shape of the magnetisation curves otherwise unchanged (data not shown).
\section{Conclusions}
In summary, we have studied how a term $\mathcal{H}'$ caused by antisymmetric
spin interactions and/or a staggered $g$ tensor affects BEC in spin-gap
systems. This interaction directly mixes the singlet with the triplet soft
mode so that its effect is non-perturbative at magnetic fields $H\sim H_c$.
Treating the magnon-magnon interaction in the one-loop approximation we have
found a gapped quasiparticle spectrum compared to the gapless spectrum without
$\mathcal{H}'$ and a nonzero condensate density for all temperatures
consistent with findings in ref.~\cite{MitraHalperin} for Haldane spin chains.
Differences to the case of crystal field anisotropies have also been
discussed. We have then performed a detailed quantitative study of the field
induced magnetisation process in TlCuCl$_3$ and found that the scattering
amplitude $v_0$ seems to be much larger than expected from magnon-magnon
interactions alone. We see this as a further confirmation that spin-phonon
coupling is important in this compound. Finally we have pointed out that even
in TlCuCl$_3$ where $\mathcal{H}'$ is expected to be tiny due to symmetry it
seems to remain crucial to obtain a quantitative correct description of the
experimental data.

\acknowledgments The authors acknowledge valuable discussions with
G.~Khaliullin, V.~N.~Kotov and F.~Mila. This work has been supported by the
Australian Research Council (ARC). JS acknowledges financial support by the
German Research Council (DFG).


\begin{thebibliography}{10}

\bibitem{AffleckBEC}
{\sc Affleck I.}, {\em Phys. Rev. B\/}, {\bf 43} (1990) 3215.
\relax
\bibitem{GiamarchiTsvelik}
{\sc Giamarchi T.} and {\sc Tsvelik A. M.}, {\em Phys. Rev. B\/}, {\bf 59}
  (1998) 11398.
\relax
\bibitem{NikuniOshikawa}
{\sc Nikuni T.}, {\sc Oshikawa M.}, {\sc Oosawa A.}, and {\sc Tanaka H.}, {\em
  Phys. Rev. Lett.\/}, {\bf 84} (2000) 5868.
\relax
\bibitem{RiceScience}
{\sc Rice T. M.}, {\em Science\/}, {\bf 298} (2002) 760.
\relax
\bibitem{FetterWalecka}
{\sc Walecka J. D.} and {\sc Fetter A. L.}, {\em Quantum theory of
  many-particle systems\/} (McGraw Hill College Div., 1971).
\relax
\bibitem{GriffinRep}
{\sc for a review~see: H.~Shi} and {\sc Griffin A.}, {\em Phys. Reports\/},
  {\bf 304} (1998) 1.
\relax
\bibitem{MatsumotoNormandPRL}
{\sc Matsumoto M.}, {\sc Normand B.}, {\sc Rice T. M.}, and {\sc Sigrist M.},
  {\em Phys. Rev. Lett.\/}, {\bf 89} (2002) 077203.
\relax
\bibitem{CavadiniHeigold}
{\sc Cavadini N.}, {\em et~al.\/}, {\em Phys. Rev. B\/}, {\bf 63} (2001)
  172414.
\relax
\bibitem{ChoiGuentherodt}
{\sc Choi K. Y.}, {\em et~al.\/}, {\em Phys. Rev. B\/}, {\bf 68} (2003)
  174412.
\relax
\bibitem{ShermanLemmens}
{\sc Sherman E. Y.}, {\em et~al.\/}, {\em Phys. Rev. Lett.\/}, {\bf 91} (2003)
  057201.
\relax
\bibitem{RueggNature}
{\sc R\"uegg C.}, {\em et~al.\/}, {\em Nature\/}, {\bf 423} (2003) 62.
\relax
\bibitem{MitraHalperin}
{\sc Mitra P. P.} and {\sc Halperin B. I.}, {\em Phys. Rev. Lett.\/}, {\bf 72}
  (1994) 912.
\relax
\bibitem{ChubukovSachdevBhatt} {\sc Chubukov A. V.}, {\em JETP Lett.\/}, {\bf
    49} (1989);  {\sc Sachdev S.} and {\sc Bhatt R.}, {\em Phys. Rev. B\/},
  {\bf 41} (1990) 9323.  
\relax
\bibitem{KotovSushkov}
{\sc Kotov V. N.}, {\sc Sushkov O.}, {\sc Weihong Z.}, and {\sc Oitmaa J.},
  {\em Phys. Rev. Lett.\/}, {\bf 80} (1998) 5790.
\relax
\bibitem{OshikawaAffleck}
{\sc Oshikawa M.} and {\sc Affleck I.}, {\em Phys. Rev. Lett.\/}, {\bf 79}
  (1997) 2883.
\relax
\bibitem{Details}
Details will be published elsewhere.
\relax
\bibitem{GlazkovSmirnov}
{\sc Glazkov V. N.}, {\sc Smirnov A. I.}, {\sc Tanaka H.}, and {\sc Oosawa A.},
  {\em Phys. Rev. B\/}, {\bf 69}  (2004) 184410.
\relax
\bibitem{Holstein}
{\sc Holstein T.}, {\em Ann. Phys. (NY)\/}, {\bf 8} (1959) 343.
\relax
\bibitem{Moriya}
{\sc Moriya T.}, {\em Phys. Rev.\/}, {\bf 120} (1960) 91.
\relax
\end{thebibliography}
\end{document}